\documentclass[a4paper, amsfonts, amssymb, amsmath, reprint, showkeys, nofootinbib, twoside]{revtex4-1}
\usepackage[english]{babel}
\usepackage[utf8]{inputenc}
\usepackage[colorinlistoftodos, color=green!40, prependcaption]{todonotes}
\usepackage{amsthm}
\usepackage{mathtools}
\usepackage{physics}
\usepackage{xcolor}
\usepackage{graphicx}
\usepackage[left=23mm,right=13mm,top=35mm,columnsep=15pt]{geometry} 
\usepackage{adjustbox}
\usepackage{placeins}
\usepackage[T1]{fontenc}
\usepackage{lipsum}
\usepackage{csquotes}
\usepackage[pdftex, pdftitle={Article}, pdfauthor={Author}]{hyperref} 
\usepackage{physics}
\usepackage{bm}
\usepackage{multirow}
\usepackage{array}
\newcolumntype{P}[1]{>{\centering\arraybackslash}p{#1}}
\newcolumntype{M}[1]{>{\centering\arraybackslash}m{#1}}

\begin{document}

\title{Physics-augmented deep learning for high-speed electromagnetic simulation and optimization}

\author{Mingkun Chen$^{1}$}
\author{Robert Lupoiu$^{1}$}
\author{Chenkai Mao$^{1}$}
\author{Der-Han Huang$^{1}$}
\author{Jiaqi Jiang$^1$}
\author{Philippe Lalanne$^2$}
\author{Jonathan A. Fan$^{1\ast}$}

\address{
1.Department of Electrical Engineering, Stanford University, Stanford, CA, USA \\ 
2.Laboratoire Photonique, Institut d’Optique Graduate School, Université Bordeaux, Bordeaux, France}

\author{
$^{\ast}$Corresponding author: \emph{jonfan@stanford.edu} \\
These authors contributed equally: Mingkun Chen, Robert Lupoiu, Chenkai Mao, Der-Han Huang.
}

\maketitle

\section{Abstract}
The calculation of electromagnetic field distributions within structured media is central to the optimization and validation of photonic devices.  We introduce WaveY-Net, a hybrid data- and physics-augmented convolutional neural network that can predict electromagnetic field distributions with ultra fast speeds and high accuracy for entire classes of dielectric photonic structures.  This accuracy is achieved by training the neural network to learn only the magnetic near-field distributions of a system and to use a discrete formalism of Maxwell's equations in two ways: as physical constraints in the loss function and as a means to calculate the electric fields from the magnetic fields.  As a model system, we construct a surrogate simulator for periodic silicon nanostructure arrays and show that the high speed simulator can be directly and effectively used in the local and global freeform optimization of metagratings.  We anticipate that physics-augmented networks will serve as a viable Maxwell simulator replacement for many classes of photonic systems, transforming the way they are designed.


\section{Introduction}
Maxwell simulators are essential tools for the characterization and design of electromagnetic systems.  These systems operate at frequencies spanning the radio wave to X-ray and include a diversity of antennas \cite{koziel2014antenna, hannan1965simulation, sun2013large}, diffractive surfaces\cite{yu2014flat, sell2017large, fan2020freeform, phan2019high, zhou2021inverse}, metamaterials \cite{gansel2009gold, valentine2008three,chen2006active}, and guided wave-based photonic circuits \cite{fan2001waveguide, piggott2015inverse,roeloffzen2013silicon}.  Amongst the most popular frequency domain Maxwell solvers are the finite element method (FEM) \cite{jin2015finite,silvester1996finite} and finite difference frequency domain (FDFD) algorithms \cite{lalanne2000numerical,ceviche, alkan2013double,rumpf2012simple}.  In both algorithms, the system domain is subdivided into discrete voxels and the simulator solves for electromagnetic fields by constructing and inverting a sparse matrix with dimensions proportional to the total number of voxels.  While these methods can be used to accurately solve general electromagnetics problems, the time and computation cost of matrix inversion serve as practical bottlenecks for the simulation of large domains and for design tasks, where large numbers of electromagnetic simulations are required to perform iterative optimization.

To address this bottleneck, deep neural networks serving as high speed surrogate Maxwell simulators have emerged as promising  algorithms that can operate orders-of-magnitude faster than conventional Maxwell simulators \cite{jiang2020deep, so2020deep, ma2021deep, an2020deep, yeung2021multiplexed, lin2021end}.  Many initial attempts to use neural networks in this manner were based on the training of fully connected deep neural networks, which could accurately model the spectral response of photonic structures described by a handful of geometric parameters \cite{peurifoy2018nanophotonic,malkiel2018plasmonic, ma2018deep,liu2018training, an2019deep}.  To model high dimensional data forms, such as electromagnetic field distributions within a photonic structure, modifications to these deep learning approaches have been proposed.  In one example, dimensionality-reduced forms of the fields were trained in conjunction with a fully connected deep network to map metasurface geometry to field distribution \cite{zhelyeznyakov2021deep}.  In another example, convolutional neural networks (CNNs) were shown to be effective at predicting electromagnetic field distribution maps within a photonic structure \cite{wiecha2019deep}.  While these demonstrations pointed to the potential of neural networks as simulators, they were exclusively trained using data with no knowledge of Maxwell’s equations \cite{Maxwell}, placing limits on their ability to process and learn wavelike electromagnetic phenomena.


To expand the capabilities of deep neural networks for physical science problems, physics-informed neural networks have been proposed that explicitly incorporate physical constraints, such as a governing differential equation, into the loss function \cite{raissi2019physics,goswami2020transfer,meng2020ppinn,kharazmi2019variational, jagtap2020adaptive}.  These concepts have been developed in the field of fluid mechanics and have been adapted to fully connected and CNN architectures \cite{raissi2018hidden, zhu2019physics}.  More recently, they have been utilized to solve vectorial electromagnetics problems, where an analytic form of Maxwell's equations and boundary conditions were used to constrain the loss function \cite{chen2020physics, lu2021physics}.  The explicit incorporation of physics into the network training process produced impressive simulation and inverse design demonstrations.  However, these methods that solve differential equation problems through network training are resource intensive to run, and they involve the solving of individual problems and not classes of problems.   


\begin{figure*}[ht]
  \centering
  \includegraphics[width=\textwidth]{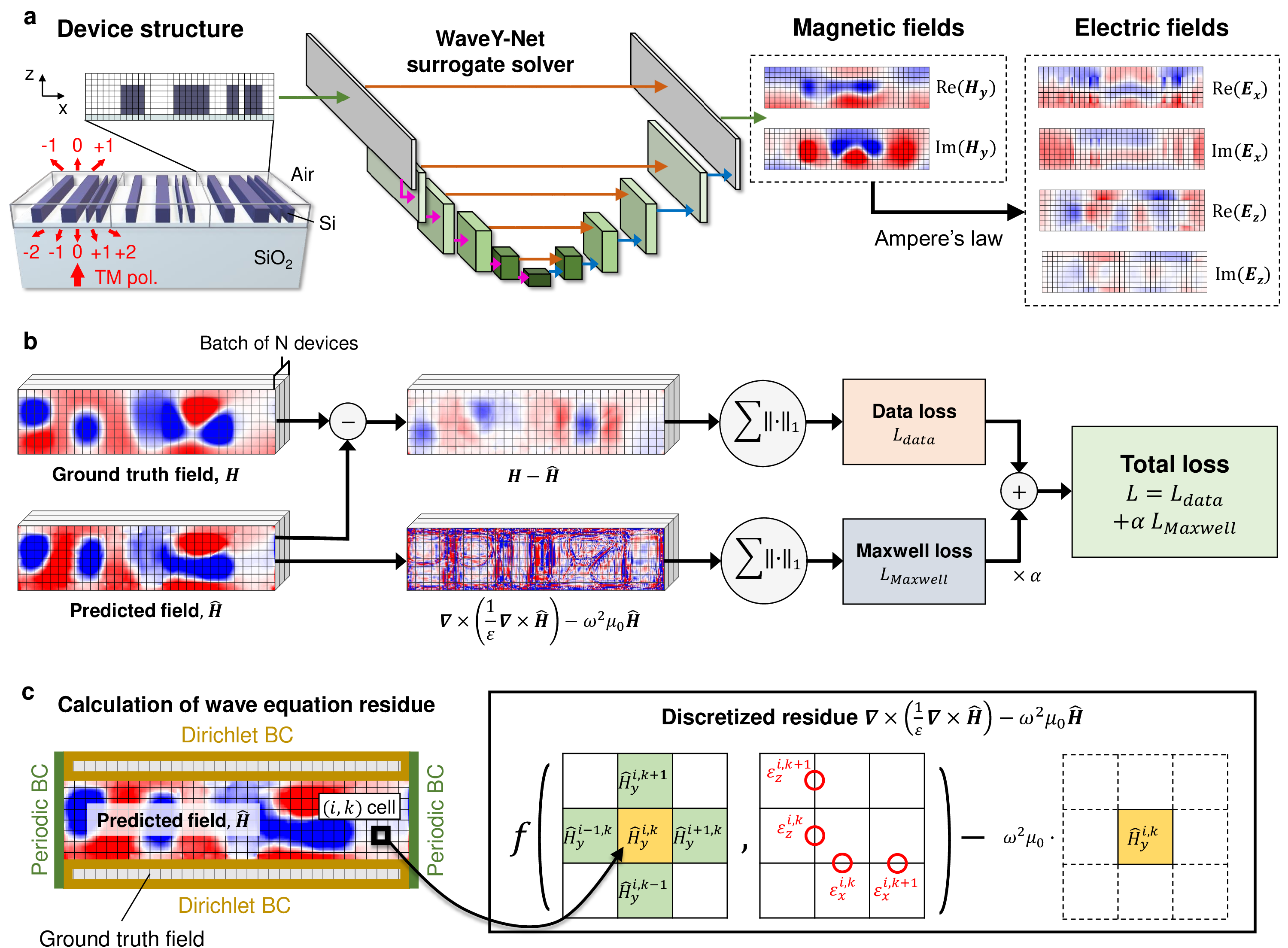}
  \caption{Overview of the WaveY-Net network architecture and training procedure. a) WaveY-Net is a UNet that learns the electromagnetic behavior of silicon-based metagratings (left).  The input is an image of a single unit cell of the metagrating, and the outputs are magnetic field maps (center).  Electric field maps are calculated from the magnetic field maps using a discrete version of Ampere's law (right).  Orange arrows: shortcut connections.  Pink arrows: periodic convolution and maxpooling operations.  Blue arrows: periodic convolution and upsampling operations.  b) Computational graph of the loss function.  The loss function comprises two terms: data loss, which quantifies deviations between ground truth and predicted magnetic field maps, and Maxwell loss, which quantifies the deviation of the predicted magnetic field maps from Maxwell's equations.  c) Calculation of Maxwell loss.  Periodic boundary conditions along vertical boundaries and Dirichlet boundary conditions along horizontal boundaries enforce a well-posed loss expression.  Maxwell loss within the magnetic field maps is calculated using a discretized version of the wave equation residue, which imposes relationships between magnetic field values at a given voxel and those of its nearest neighbors.
  }
  \label{fig:fig1}
\end{figure*}

In this Article, we introduce Maxwell surrogate simulators based on a hybrid data- and physics-augmented training approach. We term these networks WaveY-Nets, and they operate with high speed, accuracy, and generalizability, as they combine the generalization and high speed solving capabilities of data-only training with explicit Maxwell equation constraints from physics-informed training.  To maximize network accuracy and capacity, we train the networks to output the magnetic field maps of a system and use Ampere’s law to explicitly calculate the electric field maps.  This approach follows FEM and FDFD formalisms, which solve for one field type and use Maxwell's equations to calculate the other field type \cite{jin2015finite, lalanne2000numerical, rumpf2012simple}, and it takes advantage of the fact that the electric and magnetic field distributions have the same information content and do not need to be independently learned.  As a model system, we consider a two-dimensional surrogate simulator that can output field profiles internal to a periodic array of dielectric nanoridges, given a normally incident, TM-polarized plane wave.  We quantify the accuracy of these surrogate simulators and show they can be directly used in local and global adjoint optimization algorithms for designing metagratings that selectively diffract incident light to a single diffraction order.  The networks offer a two to four order of magnitude speed up depending on the optimization method.  





\section{Results}

\subsection{Network architecture and loss formulation}
The diffractive system captured by our surrogate model consists of silicon nanoridges illuminated by a normally incident beam with a wavelength of 1050 nm (Fig. \ref{fig:fig1}a).  The wave has transverse magnetic (TM) polarization, such that the electromagnetic field distribution is fully described by field components $H_y$, $E_x$, and $E_z$.  The silicon nanoridges are situated on a silicon dioxide substrate and have a period of 1600 nm, producing a total of eight diffraction orders.  The grating period contains four total silicon ridges and the ridge height is fixed to be 325 nm.  The full simulation window is defined on a 256 x 64 pixel grid and includes the silicon device and thin substrate and superstrate regions below and above the device, respectively.  This discrete representation of the device region sets the spatial resolution of the device layout and corresponding electromagnetic fields to be 6.25 nm along the horizontal and vertical directions.  The device landscape contains all devices in which the silicon ridges and air spacers have widths of 62.5 nm or greater, which corresponds to approximately three hundred billion unique device configurations.

Our network scheme is outlined in Fig. \ref{fig:fig1}a and Fig. S1 and is based on the UNet \cite{UNet}, which is a CNN architecture where the input and output data structures are images with the same dimension.  Shortcut connections strengthen relationships between the input and output images.  UNets were initially popularized in the computer vision community for image processing tasks such as image segmentation \cite{UNet, li2018h, weng2019unet, zeng2019ric}, and they are particularly well suited for our application because there exists a strong correspondence between input geometry and output field distribution.  These networks were the basis for the data-driven surrogate simulators.  The network input is a 256 x 64 pixel image of the simulation window, where the input image pixel values are normalized refractive index values with 1 and 0 corresponding to silicon and air, respectively.  The output is 256 x 64 pixel images of the real and imaginary $H_y$ field maps within the structure.  The complex $E_x$ and $E_z$ field maps are calculated from $H_y$ using Ampere’s law.  

We name our network WaveY-Net because it uses Maxwell's equations based on the Yee grid formalism to enforce wavelike field behavior of the output.  The Yee grid is an established framework for finite-difference time and frequency domain electromagnetic simulations, and it is formulated to rigorously specify spatial relationships between discrete field components, boundary conditions at dielectric discontinuities, and discrete derivative operations \cite{Yee}.  A schematic of the two-dimensional TM Yee grid is in Fig. S3.  To summarize, the $H_y$ fields are placed at the center of each pixel and parallel electric field components are placed at the pixel boundaries.  Permittivity profiles are calculated for each field component to account for the pixel-level spatial offset defining each component.  With the Yee grid, the discrete formulation of Ampere's law used to calculate $E_x$ and $E_z$ from $H_y$ are as follows: 
\begin{equation}\label{ampereslaw1}
    \frac{{H_y}^{i,k}-{H_y}^{i,k-1}}{\Delta z} = -i \omega {\varepsilon_{x}}^{i,k}{E_x}^{i,k}
\end{equation}

\begin{equation}\label{ampereslaw2}
    \frac{{H_y}^{i,k}-{H_y}^{i-1,k}}{\Delta x} = i \omega {\varepsilon_{z}}^{i,k}{E_z}^{i,k}
\end{equation}
$i$ and $k$ are discrete index labels for the horizontal and vertical pixel positions, respectively.

To train WaveY-Net, the loss function is specified to have two components (Fig. \ref{fig:fig1}b):
\begin{equation}\label{totloss}
    L = L_{data} + \alpha L_{Maxwell}
\end{equation}
$L_{data}$ is data-driven loss from which the network attempts to fit the network output fields with a ground truth training set.  It takes the form of mean absolute error (MAE): 
\begin{equation}\label{dataloss}
    L_{data} = \frac{1}{N}\sum_{n=1}^{N} \lVert \emph{\textbf{H}}^{(n)}-\widehat{\emph{\textbf{H}}}^{(n)} \lVert_{1}
\end{equation}
$\emph{\textbf{H}}$ represents ground truth field profiles from the training set, $\widehat{\emph{\textbf{H}}}$ are the field profiles outputted from the network, $N$ is a given batch size, and $n$ is the index of the device within a batch.  $L_{Maxwell}$ specifies the compliance of the outputted fields with Maxwell's equations and is the MAE of the magnetic field wave equation residue:
\begin{equation}\label{maxwellloss}
    L_{Maxwell} = \frac{1}{N}\sum_{n=1}^{N}  \lVert
    \nabla \times (\frac{1}{\bm{\varepsilon}^{(n)}} \nabla \times \widehat{\emph{\textbf{H}}}^{(n)})
    - \omega^2 \mu_0  \widehat{\emph{\textbf{H}}}^{(n)}
    \lVert_{1}
\end{equation}
The discrete Yee grid-based formalism of the magnetic field wave equation is Equation S1 in the Supplementary Section 2.  If the calculated Maxwell loss for a given pixel in an outputted field profile is zero, it means that the fields local to that pixel are consistent with Maxwell's equations. $\alpha$ is a hyperparameter that balances the contributions of data and Maxwell loss and it is dynamically tuned during network training in a manner that stabilizes the training process.  More details pertaining to the network architecture and training methodology are in the Method Section and Supplementary Sections 1 and 3.


\begin{figure*}[ht]
  \centering
  \includegraphics[width=\textwidth]{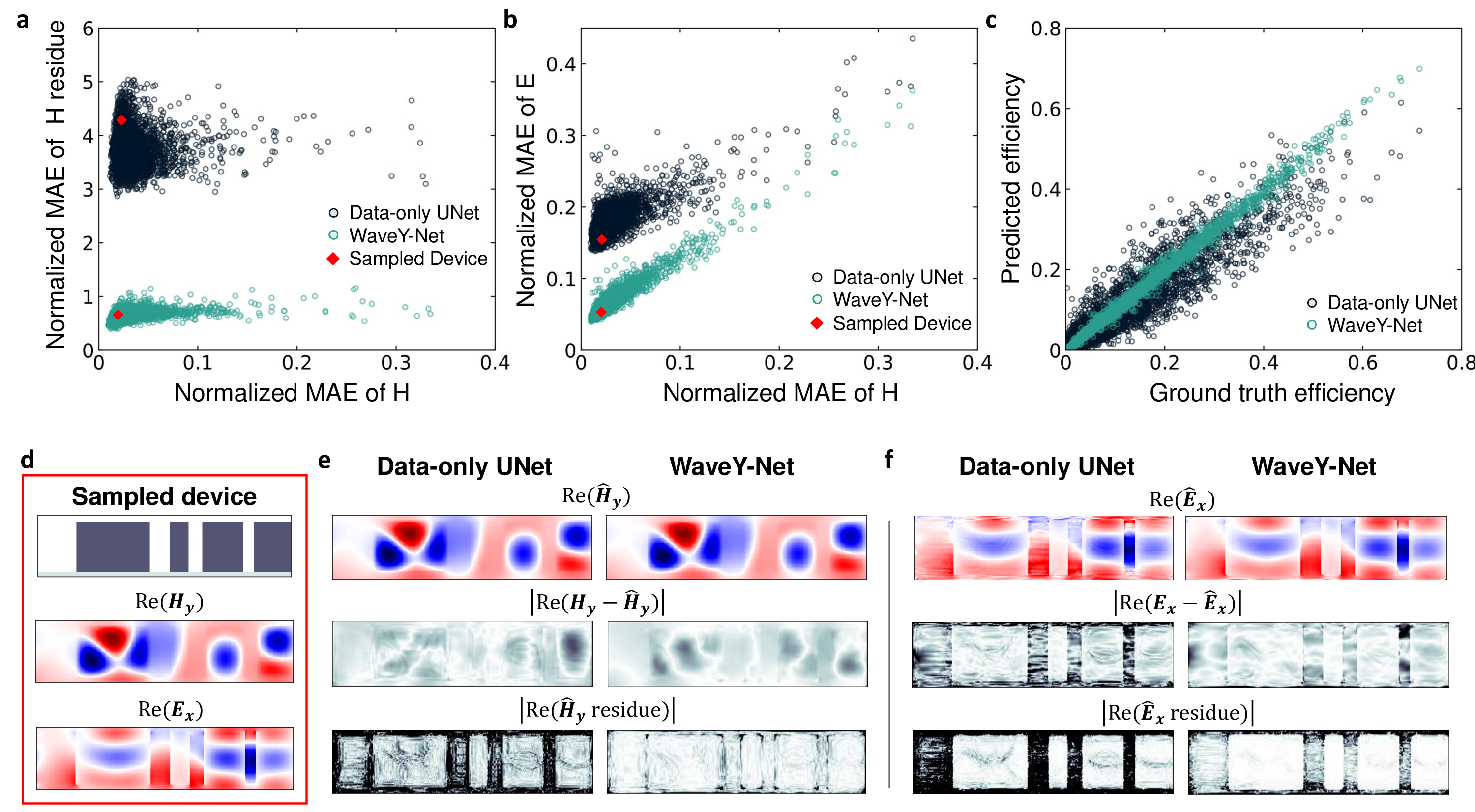}
  \caption{Benchmark comparison between WaveY-Net and a data-only UNet, both which predict magnetic field maps. a) Scatter plot of magnetic wave equation residue MAE versus magnetic field MAE.  The predicted magnetic field maps are from 3000 test devices evaluated by each network.  b) Scatter plot of the electric field MAE versus magnetic field MAE for the same devices. The definition of normalized MAE is in Method Section.  c) Scatter plot of the predicted versus ground truth far-field diffraction efficiencies for the same devices, for light diffracting into the transmitted $+1$ order.  d) Device layout, $H_y$ field map, and $E_x$ field map of a representative, randomly sampled device from the test set.  e) Magnetic and f) electric field maps, field MAE maps, and wave equation residue MAE maps for the sampled device evaluated by both networks. H residue: wave equation residue of the magnetic field. E residue: wave equation residue of the electric field. Color scales: H-field [-0.015, 0.015], H-field MAE [0, 0.001], H residue [0, 0.01], E-field [-3, 3], E-field MAE [0, 0.2], E residue [0, 20].
  }
  \label{fig:fig2}
\end{figure*}

While $L_{data}$ and $L_{Maxwell}$ both reduce to zero as the outputted fields converge to ground truth values, each loss term captures different information about the fields.  $L_{data}$ captures the accuracy of the fields on a pixel-by-pixel basis, ensuring that the outputted fields are close to ground truth values but without explicitly accounting for spatial relationships between pixels.  $L_{Maxwell}$, on the other hand, utilizes discrete spatial derivatives that explicitly capture physical relationships between neighboring field pixels.  These spatial correlations can be visualized by the calculation of $L_{Maxwell}$ for a given pixel, shown in Fig. \ref{fig:fig1}c, which has the form of a 3 x 3 pixel kernel.  As a kernel, $L_{Maxwell}$ can be efficiently calculated for every pixel in a field map using a graphical processing unit (GPU), which naturally processes convolution operations in a fast and parallel manner.  With the complementary nature of these loss functions, it is possible for $L_{data}$ to be low and $L_{Maxwell}$ to be high if the outputted field values are approximately correct but there are large field fluctuations between neighboring pixels.  It is also possible for $L_{data}$ to be high and $L_{Maxwell}$ to be low if the local fields are wavelike but dissimilar to the ground truth values.  

To ensure that $L_{Maxwell}$ is well posed, it is essential that the loss function incorporates proper boundary conditions.  Without proper boundary conditions, $L_{Maxwell}$ can push the outputted fields to wavelike profiles that locally satisfy Maxwell's equations but are far from ground truth values.  At the left- and right-most columns of the field maps, we naturally apply periodic boundary conditions within the UNet framework by applying periodic padding to the convolutional calculations along the x-axis (Fig. \ref{fig:fig1}c). Dirichlet boundary conditions at the horizontal field map boundaries are enforced by substituting the top and bottom rows of the predicted field maps with ground truth magnetic field values (see a more detailed explanation in the Supplementary Section 4). In this manner, $L_{Maxwell}$ is zero at all pixels only when the fields converge exactly to ground truth values.

\subsection{WaveY-Net solver}

To evaluate the impact of $L_{Maxwell}$ on UNet simulator accuracy, we train WaveY-Net, which trains using both $L_{data}$ and $L_{Maxwell}$, and compare it with a data-only UNet trained with only $L_{data}$.  In both cases, the neural networks use 30,000 random device layouts and their associated fields as training data, have outputs consisting of the real and imaginary magnetic field maps, and use Equations \ref{ampereslaw1} and \ref{ampereslaw2} to calculate the electric field maps.  Training data are generated using an open source FDFD solver \cite{ceviche}.  A summary of the performance of both networks, compiled from 3,000 test data, is presented as scatter plots in Figs. \ref{fig:fig2}a-\ref{fig:fig2}c and Table \ref{tab:table1}.  We find that WaveY-Net and the data-only UNet are reasonably accurate magnetic field surrogate solvers, with magnetic field MAE averages of 0.033 and 0.036, respectively (Fig. \ref{fig:fig2}a).  As such, the addition of $L_{Maxwell}$ produces a modest but not substantial improvement in predicted magnetic field accuracy, as defined on a pixel-by-pixel basis.

However, the fields outputted by WaveY-Net are more self-consistent with the magnetic field wave equation compared to the data-only UNet, with an approximately six times difference in $L_{Maxwell}$ MAE between the two networks (Fig. \ref{fig:fig2}b).  This consistency of the magnetic field maps with the magnetic field wave equation leads to more accurate calculations of the electric fields, with WaveY-Net producing electric fields with average MAE values over twice smaller than those from the data-only UNet and with significantly improved consistency with the electric field wave equation (Table \ref{tab:table1}).



To further quantify the utility of $L_{Maxwell}$, we perform near-to-far-field transformations \cite{snyder2012optical} on the electric fields produced by each network and calculate diffraction efficiencies into the transmitted $+1$ order.  Accurate far-field amplitudes and phases are required for tasks such as local and global freeform optimization and will be utilized later in this study.   More details pertaining to the near-to-far-field calculation are in the Method Section and Supplementary Section 5.  Scatter plots of the ground truth and predicted diffraction efficiencies for fields generated by the two networks are presented in Fig. \ref{fig:fig2}c.  For field plots generated by the data-only UNet, there are clear deviations between the predicted and ground truth diffraction efficiencies, with an average efficiency error of 3.7\%.  WaveY-Net, on the other hand, outputs fields that produce relatively accurate near-to-far-field efficiency calculations, with an average efficiency error of only 0.67\%.  This error does not increase for high efficiency devices in spite of the fact that there is a disproportionately low number of high efficiency devices in the network training set.  As such, Maxwell regularization is effective at predicting usual figures of merit, e.g. efficiency, by enforcing long range consistency through local field-map constrains.




To more clearly visualize the discrepancy between the fields produced by both networks, we examine the electromagnetic fields outputted by each network for a randomly selected test device (Fig. \ref{fig:fig2}d).  Ground truth $\mbox{Re(}E_x\mbox{)}$ and $\mbox{Re(}H_y\mbox{)}$ field maps are also shown, with the other field components presented in the Supplementary Section 6. The outputted magnetic field maps from both networks are in Fig. \ref{fig:fig2}e and display good agreement with the ground truth field map.  Their corresponding MAE maps show pixel-level deviations that each feature approximately 3\% relative error.  However, the MAE map from the data-only UNet has a higher non-physical spatial frequency noise component.  This difference is attributed to the spatial filtering functionality of the $L_{Maxwell}$ loss term, which enforces spatial derivative constraints between neighboring pixels and pushes the fields to have spatially smooth, wavelike forms.  The benefits featured by Maxwell regularization carry over to the calculation of electric field, where the electric field maps produced from WaveY-Net are relatively smooth and low in error while the data-only UNet electric fields have visible high spatial frequency noise and regions featuring as high as 25\% error (Fig. \ref{fig:fig2}f).  Errors are particularly amplified at dielectric discontinuities, where the electric field components have discontinuities themselves, and they carry over to the electric field wave equation residue maps.

\begin{table}
\begin{tabular}{ M{1.75cm}|M{1cm}|M{2.5cm} M{2cm} }
 \multicolumn{2}{c|}{Normalized MAE} &Data-only UNet  &  WaveY-Net \\
 \hline
  \multirow{2}{*}{$\emph{\textbf{H}}-\widehat{\emph{\textbf{H}}}$}  & Avg    &0.036&   0.033\\
 &   Std  & 0.027   &0.027\\
 \hline
 \multirow{2}{*}{$\widehat{\emph{\textbf{H}}}$ residue} &Avg & 3.774 &  0.646\\
 &Std & 0.361&  0.087\\
 \hline
 \multirow{2}{*}{$\emph{\textbf{E}}-\widehat{\emph{\textbf{E}}}$} &   Avg  & 0.182&0.071\\
 & Std  & 0.024   &0.026\\
 \hline
 \multirow{2}{*}{$\widehat{\emph{\textbf{E}}}$ residue}&   Avg &229.73 & 25.34  \\
 & Std &17.07 & 3.72    \\
\end{tabular}

\caption{\label{tab:table1} Summary of average MAE values and associated standard deviation values for the test data analyzed in Fig. \ref{fig:fig2}a.}
\end{table}

\begin{figure}[ht]
  \centering
  \includegraphics[width=0.8\linewidth]{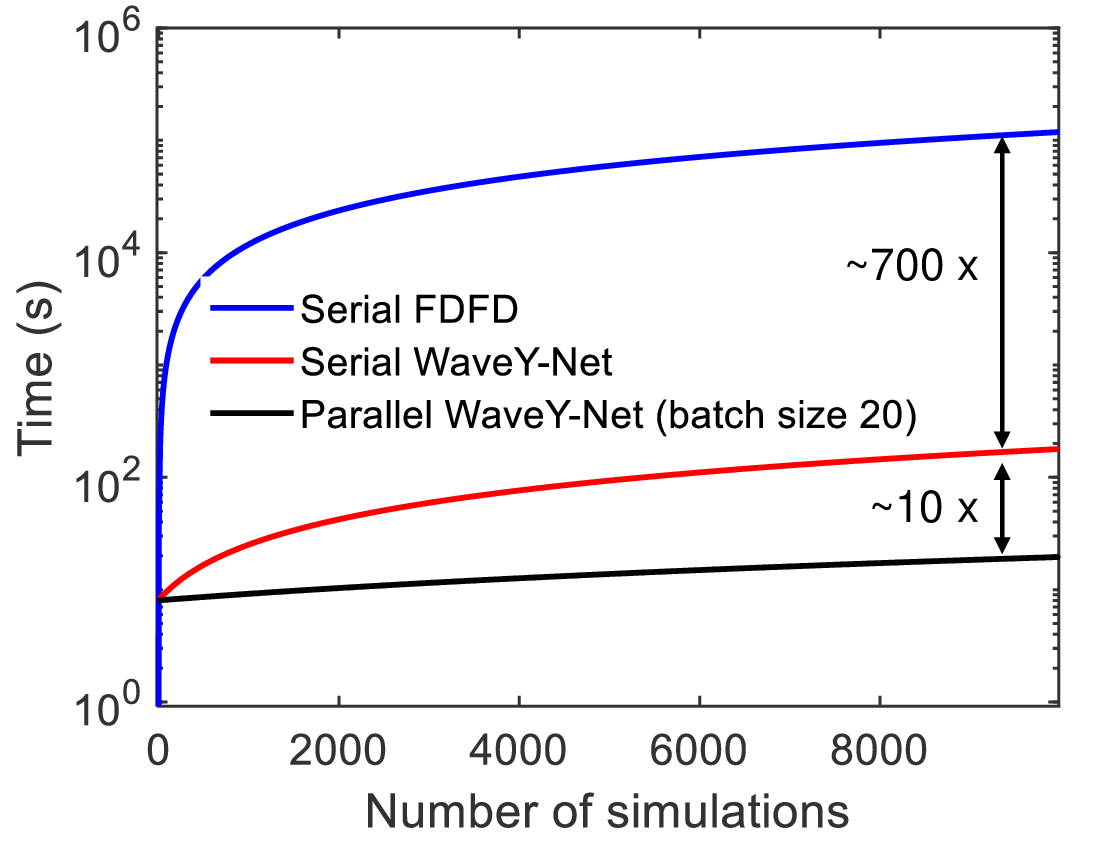}
  \caption{Plot of computation time versus number of simulations for three different simulation methods: a conventional FDFD solver; serial WaveY-Net, which simulates individual devices in series; and parallel WaveY-Net, which simulates devices in batches of twenty in parallel.
  }
  \label{fig:fig3}
\end{figure}

The acceleration in computation enabled by WaveY-Net, compared to a conventional full wave solver, is significant due to a combination of software and hardware features.  A summary of the computation time required by a conventional FDFD solver \cite{ceviche} and WaveY-Net for different numbers of simulations is shown in Fig. \ref{fig:fig3}.  A single simulation is defined as the evaluation of the complex magnetic field maps for an individual device pattern.  Computations for the conventional FDFD solver are performed with four cores of a 2.70 GHz Intel Xeon Gold 6150 CPU Processor with 32Gb of RAM, and those for WaveY-Net are performed with one NVIDIA A100 GPU with 40GB of VRAM and PCIE connection.  The serial WaveYNet solver provides a well over two orders of magnitude speedup than the FDFD solver and the parallel WaveYNet provides nearly a four orders of magnitude faster speedup in computation.  For the evaluation of 7,000 devices, the FDFD solver takes approximately one day while the serial WaveYNet, which evaluates one device at a time, take three minutes.  The parallel WaveY-Net evaluates twenty devices at a time by taking advantage of the parallel computing capability of GPUs and can evaluate 7,000 devices within 20 seconds. More detailed comparison can be found in Supplementary Section 8. It is also noted that these speedups featured by WaveY-Net lead to significant reductions in environmental impact, which is further detailed in the Supplementary Section 11.

\subsection{WaveY-Net-based freeform optimizers}

We next turn our attention to the utilization of WaveY-Net in design and optimization algorithms, which are ideal platforms for benchmarking high speed electromagnetic solvers because they require the iterative evaluation of distinct device structures and can require batches of devices to be evaluated at a given time.  Our focus will be on the freeform boundary optimization of dielectric metagratings that selectively diffract normally incident light to the transmitted $+1$ diffraction order.  We consider two methods for optimization: gradient-based local optimization based on the adjoint variables method \cite{sigmund2013topology, miller2012photonic, sell2017large, chen2020design} and global optimization based on GLOnets \cite{jiang2019simulator, jiang2019global, chen2020design}, which is a population-based evolutionary algorithm that performs optimization through the training of a generative neural network. 

\begin{figure*}[ht]
  \centering
  \includegraphics[width=\textwidth]{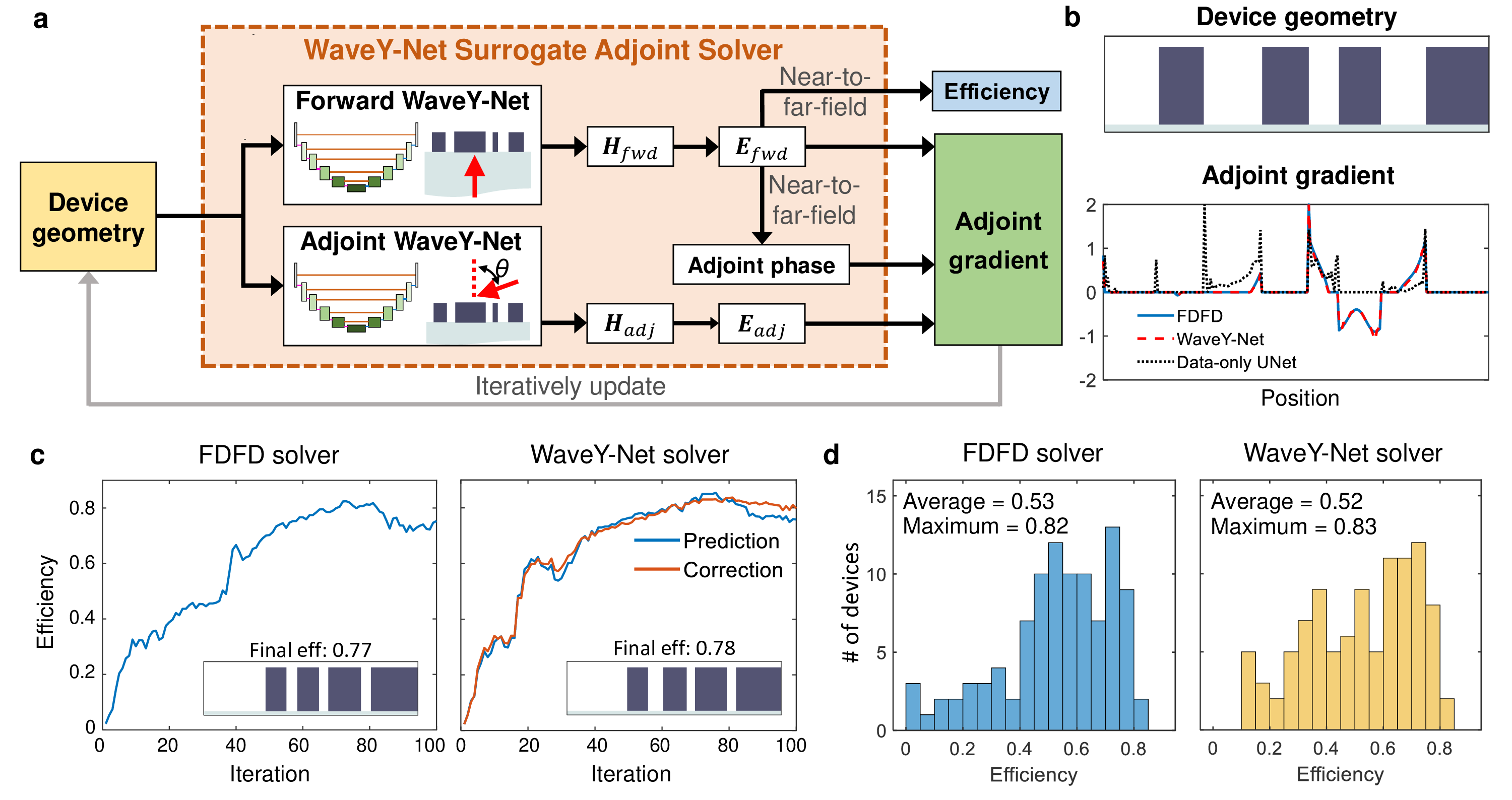}
  \caption{Local freeform metagrating optimization based on a WaveY-Net surrogate adjoint solver. a) Computational graph of the local adjoint optimizer that maximizes device diffraction efficiency into the transmitted $+1$ diffraction order.  Two WaveY-Nets are used to perform forward and adjoint magnetic field simulations, which are used to calculate diffraction efficiency and the adjoint gradient. b) A representative, randomly sampled device layout and the corresponding adjoint gradient calculated using three different electromagnetic solvers: an FDFD solver that produces ground truth gradients, WaveY-Net, and a data-only UNet.  The gradients calculated using the FDFD solver and WaveY-Net are nearly identical. c) Optimization trajectories for local adjoint optimizations performed using an FDFD solver and WaveY-Net.  The blue efficiency curve in the WaveY-Net plot is predicted by WaveY-Net while the orange efficiency curve is based on FDFD simulations. The insets show the final device layouts and efficiencies.  d) Histograms of 100 locally optimized devices with calculations performed using the FDFD solver and WaveY-Net.
  }
  \label{fig:fig5}
\end{figure*}

A computational graph of the local adjoint-based optimizer is shown in Fig. \ref{fig:fig5}a.  At the core of this algorithm is the adjoint solver, which takes the device geometry as the input and outputs the diffraction efficiency and adjoint gradient of the device.  The gradient is used to iteratively perturb the device geometry in a manner that improves device diffraction efficiency.  We formulate a WaveY-Net surrogate adjoint solver that utilizes two separately-trained WaveY-Nets.  The first is a 'forward' simulator that predicts the magnetic fields $\emph{\textbf{H}}_{fwd}$ given a normally incident plane wave.  The second is an 'adjoint' simulator that predicts the magnetic fields $\emph{\textbf{H}}_{adj}$ given an obliquely incident plane wave oriented oppositely to that of the transmitted diffracted beam (Fig. \ref{fig:fig5}a). The electric field maps $\emph{\textbf{E}}_{fwd}$ and $\emph{\textbf{E}}_{adj}$ are calculated from each corresponding magnetic field map using Equations \ref{ampereslaw1} and \ref{ampereslaw2}.  The calculated electric field maps from the forward simulator are also used to calculate the far-field amplitude and phase response in the desired diffraction order, using a near-to-far-field transformation, which yield diffraction efficiency and the adjoint phase.  The adjoint gradient is calculated by integrating the forward and adjoint electric fields and the adjoint phase.  More details are provided in the Method Section.  


To benchmark the adjoint solver, we calculate the adjoint gradient maps for a random device using three types of simulators: a fullwave FDFD simulator, a data-only UNet, and WaveY-Net.  The gradient maps are shown in Fig. \ref{fig:fig5}b and indicate that the gradients from the WaveY-Net match well with the ground truth values computed with the FDFD solver. On the contrary, the adjoint gradients calculated using the data-only UNets exhibit large errors in magnitude and even sign, indicating these solvers are not sufficiently accurate for use in optimization (More details in the Supplementary Section 9).  

To evaluate the efficacy of the WaveY-Net adjoint solver to perform optimization, we use this solver and the FDFD-based adjoint solver to perform full gradient optimizations.  Fig. \ref{fig:fig5}c summarizes the results for representative optimization runs where the same starting device geometry and 100 iterations are used for each solver.  The algorithm using the FDFD solver shows a nearly monotonic increase in efficiency over the course of the iterative process, with a final device exhibiting a 77\% diffraction efficiency.  The optimizer based on WaveY-Nets shows similar behavior and produces a final device exhibiting a 78\% diffraction efficiency, indicating that the surrogate adjoint solver is sufficiently accurate to perform iterative optimization.  The final efficiency value cited here is calculated using a fullwave solver.  Ground truth device efficiencies at each optimization iteration (orange curve) nearly match predicted values (blue curve), indicating that our near-to-far-field calculations of WaveY-Net-outputted fields yield consistently accurate diffraction efficiencies.  The optimization curves do not match exactly, indicating that small amounts of error in the gradient and efficiency calculations produce small deviations in the optimization trajectory.  To further benchmark the consistency of our WaveY-Net adjoint solver, we perform 100 local optimizations with random starting devices using WaveY-Net and the FDFD simulator.  Final device efficiency histograms for optimizers using each simulator are  summarized in Fig. \ref{fig:fig5}d and show both distributions to have similar average and maximum efficiencies.





\begin{figure*}[ht]
  \centering
  \includegraphics[width=\textwidth]{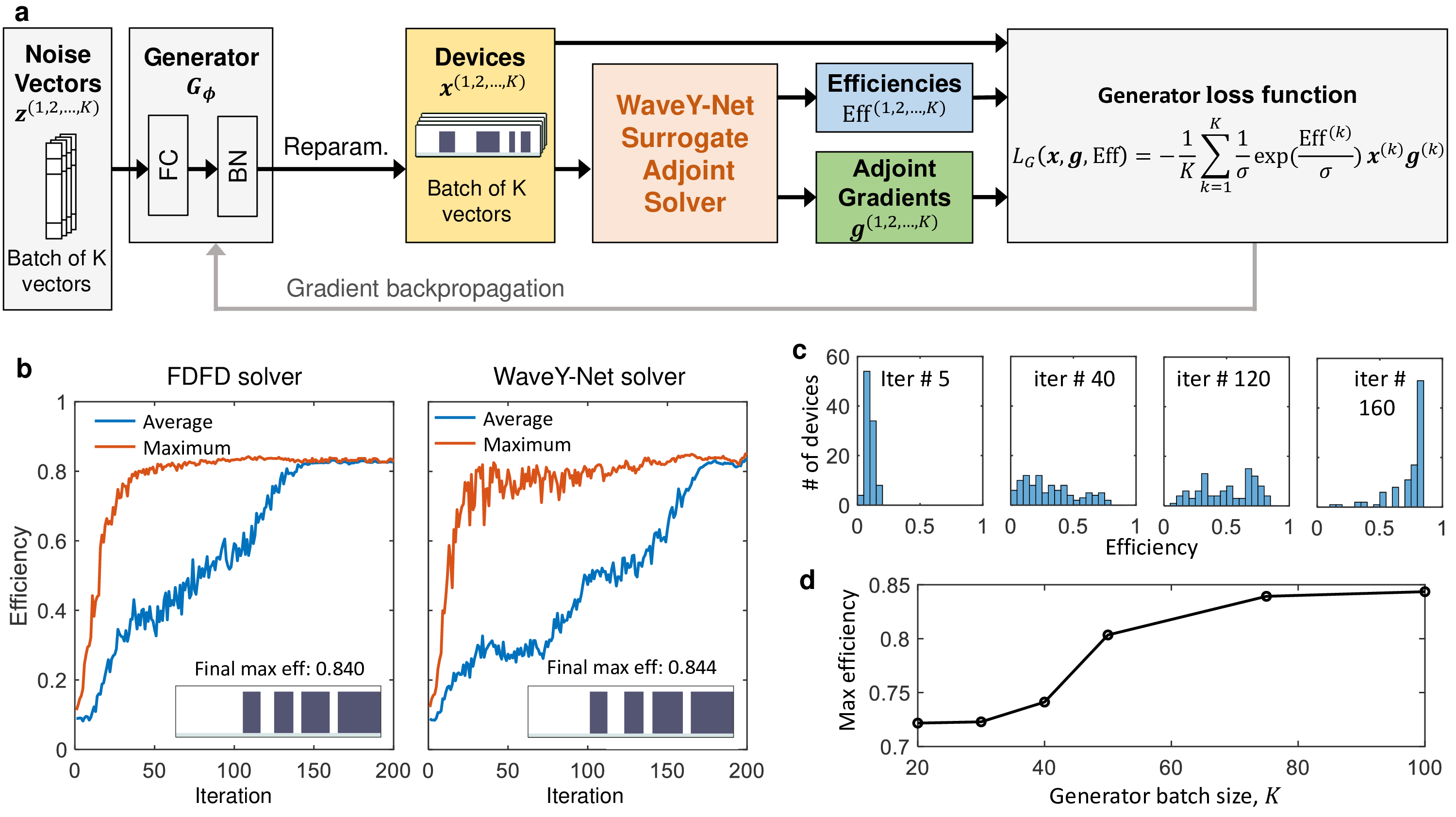}
  \caption{WaveY-Net-GLOnet algorithm for the population-based search of the global optimum. a) Computational graph of the WaveY-Net-GLOnet algorithm.  The WaveY-Net surrogate adjoint solver is the module featured in Fig. \ref{fig:fig5}a.  To perform optimization, a batch of devices with latent space representations is produced by the generator and transformed into physical devices using reparameterization.  Device efficiencies and adjoint gradients, computed by the adjoint solver, are utilized in a custom loss function to push the generated device distribution towards the global optimum. FC: fully-connected layer. BN: batch normalization layer. b) Optimization trajectories of GLOnet runs based on FDFD and WaveY-Net solvers.  The final optimal device layouts and efficiency values, shown in the insets, are similar. Batch size is $K = 100$. c) Histogram of device efficiencies as a function of iteration number from the WaveY-Net-GLOnet run. d) Plot of final optimal device efficiency versus GLOnet training batch size obtained from WaveY-Net-GLOnet.
  }
  \label{fig:fig6}
\end{figure*}

The WaveY-Net-based adjoint solver can also be directly incorporated into the GLOnets algorithm. The optimization pipeline is shown in Fig. \ref{fig:fig6}a and involves three principle parts: the iterative generation of a batch of devices from a generative network, calculating the performance gradients and efficiencies of those devices to evaluate a loss function, and  updating the network weights in the generative network with backpropagation in a manner that minimizes the loss function. The generator $G_{\phi}$ contains a single fully-connected layer followed by a batch normalization layer, where $\phi$ are the network weights, and it produces a distribution of devices $\emph{\textbf{x}}$ from uniformly distributed noise vectors  $\emph{\textbf{z}}$. To enforce a minimum feature size of 62.5 nm throughout the optimization process, we use a reparameterization transformation in which an analytic transformation converts network-outputted latent device representations to physical, constrained devices \cite{chen2020design}.  The loss function, formally derived in Ref. \cite{jiang2019simulator}, is defined to be:
\begin{equation}\label{glonetloss}
    L_G(\emph{\textbf{x}}, \emph{\textbf{g}}, \mbox{Eff}) = -\frac{1}{K}\sum_{k=1}^{K}  \frac{1}{\sigma} \exp\left(\frac{\mbox{Eff}^{(k)}}{\sigma}\right)\ \emph{\textbf{x}}^{(k)}\cdot \emph{\textbf{g}}^{(k)}
\end{equation}
The efficiencies, $\mbox{Eff}$, and adjoint gradients, $\emph{\textbf{g}}$, of the devices are calculated with the WaveY-Net adjoint solver.  As the WaveY-Net adjoint solver relies on GPU hardware, it can evaluate full batches of devices in parallel.  $K$ is the generator batch size and $\sigma$ is a hyperparameter for which 0.1 is used for optimal performance.



We run the WaveY-Net-GLOnet algorithm for the same metagrating example discussed above and also run an FDFD-based GLOnet as a ground truth benchmark.  200 optimization iterations, each with a generator batch size of 100 devices, are used in both cases.  The results are summarized in Fig. \ref{fig:fig6}b and show that WaveY-Net-GLOnet is able to converge to nearly the same optimal device as the FDFD-based GLOnet. The best sampled device has an efficiency of 0.84, which is higher than the best locally-optimized device in Fig. \ref{fig:fig5}d and indicative of an effective global search algorithm.  Fig. \ref{fig:fig6}c shows histograms of generated device efficiencies as a function of iteration number for the WaveY-Net-GLOnet and show that the generator initially has no knowledge of high quality metagrating designs, but over the course of training, the best generated device and the overall device distribution shift towards higher efficiency regimes. A plot of the best device efficiency obtained using WaveY-Net-GLOnet with different batch sizes (Fig. \ref{fig:fig6}d) shows that to effectively search for the global optimum, a batch size of at least 100 devices is ideal.  The requirement of large batch sizes plays to the strengths of the UNet simulator approach, where the forward and adjoint simulations of entire batches are computed in parallel.  We calculate that WaveY-Net-GLOnet runs approximately 7,000 times faster than FDFD-based GLOnet, which is consistent with the  trends shown in Fig. \ref{fig:fig3}.


\section{Discussion}

\begin{figure*}[ht]
  \centering
  \includegraphics[width=\textwidth]{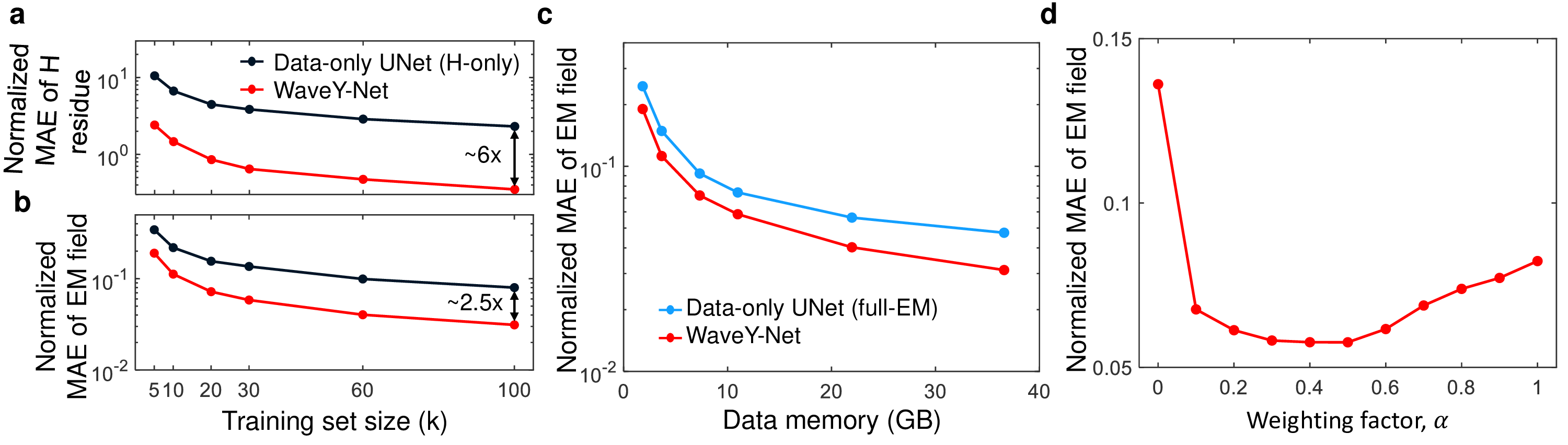}
  \caption{Benchmark numerical experiments with WaveY-Net. a) Plot of magnetic wave equation residue MAE versus training set  batch size, computed for WaveY-Net and a data-only UNet that outputs magnetic fields. b) Plot of electromagnetic field MAE versus training set batch size, computed for WaveY-Net and a data-only UNet that outputs magnetic fields. c) Plot of electromagnetic field MAE versus total training set memory size, computed for WaveY-Net and a data-only UNet that outputs the full electromagnetic fields. d) Plot of electromagnetic field MAE versus the hyperparameter $\alpha$, computed for WaveY-Net with normalized $L_{Maxwell}$ and $L_{data}$ loss terms.
  }
  \label{fig:fig4}
\end{figure*}

We perform further WaveY-Net benchmarking with two types of data-only UNets: the previously discussed two channel H-only network that outputs magnetic field maps and a six channel full-EM network that outputs field maps for all electromagnetic field components (See Supplementary Section 10 for more details).  With the H-only data-only UNets, we previously observed that the lack of Maxwell regularization led to inaccurate electric field maps.  In spite of these shortcomings, one might hypothesize that these field inaccuracies could be mitigated by simply increasing the dataset size.  This 'big data' mentality arises from trends observed in computer vision and natural language processing tasks, where it is observed that models generalize better with less overfitting when larger training sets are used \cite{barbedo2018impact, catal2009investigating, shorten2019survey, fadaee2017data}. To investigate the impact of training set size on network performance, we train data-only UNet and WaveY-Net models with the same network architecture on datasets with a total of 5k, 10k, 20k, 30k, 60k and 100k random training devices. 

Comparisons of the magnetic wave equation residue and full field MAE as a function of training set size for each model is presented in Figs. \ref{fig:fig4}a and \ref{fig:fig4}b.  We find that WaveY-Nets produce magnetic wave equation residue values that are approximately ten times lower than those from the data-only UNets, independently of the training set size.  Furthermore, the residue value from the data-only UNet trained with 100K data is similar to the residue value from WaveY-Nets trained with only 5K data (Fig. \ref{fig:fig4}a).  Without explicit Maxwell regularization, data-only UNets are not able to learn wavelike correlations between neighboring pixels, even in the limit of large training sets.  An examination of the full field MAE trends in Fig. \ref{fig:fig4}b indicates that even with 100K training data, the data-only UNet stil has limited accuracy, with normalized field MAEs of approximately 10\%.  In addition, WaveY-Net requires approximately 10x less data than the data-only UNet to produce total electromagnetic field maps with similar MAE. More discussions are in the Supplementary Section 7. This relative reduction in training data preparation is critical for our application, where the generation of training data consumes the vast majority of computational resources used for algorithm development. 


A comparison of WaveY-Net with a full-EM data-only UNet shows that WaveY-Net has a performance advantage.  When both networks are trained with 30,000 training devices, the normalized MAE of the full fields (see Method Section), is $0.054$ from WaveY-Net while the MAE values from the data-only network is $0.056$.   This slight advantage exists in spite of the fact that WaveY-Net uses nearly three times less total training data, as quantified by data memory.  The origin of this advantage is clarified when examining the normalized MAE of the network-outputted magnetic fields, which is $0.033$ for WaveY-Net and $0.046$ for the full-EM data-only UNet.  This relatively large difference arises because WaveY-Net learns data relationships for a lower dimensional problem (i.e., two output channels instead of six), allowing the trained network to display improved accuracy and generalizability.  These trends suggest that the full WaveY-Net fields can be further improved, in a manner that far surpasses the capabilities of the full-EM data-only UNet, by improving the method for calculating electric fields from magnetic fields.  

A fairer comparison, from the point of view of network training, is to benchmark both networks using the same amount of training data as quantified by data memory.  Practical limits to maximum usable training set memory are bounded by hardware limitations such as disk capacity and GPU memory.  In addition, the speed and computational resources used for network training, which involve data processing in CPUs and data transfering between GPUs and CPUs, directly correlate with training set memory. The results are summarized in Fig. \ref{fig:fig4}c and show a clear performance improvement with WaveY-Net.  It indicates that for a fixed amount of training data memory, it is always advantageous to have a more diverse dataset with less information contained by each data point and to use physical relationships to compute the rest of the information, instead of having the network learn all the information with a reduced dataset size.

Finally, we examine the impact of the weighting factor $\alpha$ in Equation \ref{totloss} to further elucidate the network training process and the relative roles of $L_{Maxwell}$ and $L_{data}$.  On one hand, hybrid networks trained with a strong $L_{Maxwell}$ weighting can be treated as physics-informed neural networks that predominantly train by solving differential equations but use data to help with network convergence.  On the other hand, hybrid networks that use a strong $L_{data}$ weighting can be treated more as conventional data-based networks that use Maxwell regularization to push the outputted data to be more wavelike.  We train a series of WaveY-Nets in which $L_{Maxwell}$ is normalized each iteration to have the same magnitude as $L_{data}$ and $\alpha$ is fixed to a chosen number.  The plot of the resulting full field MAE values for $\alpha$ ranging from 0 to 1 is shown in Fig. \ref{fig:fig4}d and indicates that the best performing networks use an $\alpha$ between 0.2 and 0.6.  As such, WaveY-Net most effectively operates as a data-based network that uses physics to regularize the quality of outputted fields.  This biasing towards data-based loss is reflected in our observation that while it is straight forward to effectively train a network only with $L_{data}$, the network does not properly converge when trained only with $L_{Maxwell}$ (See Supplementary Section 3).  Training methods that use stronger $L_{Maxwell}$ weighting are of interest because their proper implementation may reduce the reliance of large training datasets.  Concepts such as the incorporation of an active weighting scheme for boundary condition contributions may improve the performance of those networks \cite{wang2021learning}, and they will be a topic of future study.

In summary, we show that WaveY-Net, which trains using data and physical constraints, can serve as effective electromagnetic solvers.  These surrogate simulators can produce accurate field solutions for classes of freeform devices comprising four silicon nanoridges, and they can be directly used in local and global optimization algorithms that search within this design space.  An important feature of our approach is that it is data efficient, training with only a single field type and taking advantage of the explicit relationships between electric and magnetic fields fixed by Maxwell's equations.  This feature enhances the generalization capabilities of the network through optimal use of the network capacity.  This is a particularly important consideration when adapting UNets to large, three-dimensional systems, where the generation and utilization of large training data sets is extremely computationally intensive.  While our network considers devices with fixed topology and material type, we anticipate that ensembles of WaveY-Net solvers can be collectively used to solve more general classes of photonics problems.  We also anticipate that for applications that require accuracy convergence quantification, WaveY-Net can serve as a preconditioner for a rigorous Maxwell solver, providing a compromise between acceleration and accuracy \cite{trivedi2019data}.


\section{Methods}

\subsection{Network architecture}
WaveY-Net is implemented by using a traditional encoder-decoder UNet architecture consisted of six successive residual blocks, with each residual block containing six periodic-convolutional layers followed by batch normalization and a leaky rectifying linear unit (leaky ReLU). The periodic-convolutional layers uses zero padding for horizontal boundaries and pads vertical boundaries using columns from the opposite side to account for periodic boundary condition. The number of convolutional kernels doubles after each pooling layer in the encoder, which is mirrored in the decoder. Two residual connections each across three convolutional layers are implemented within the residual block, which are proven to be beneficial for efficient optimization, as well as higher accuracy especially for deep networks\cite{UNet}. For the first two encoding blocks, non-uniform maxpooling is used where the window size is $(1 \times 2)$. Correspondingly, for the last two decoding blocks, non-uniform Upsampling is implemented. Shortcut connections are utilized between corresponding encoding and decoding blocks such that the last leaky ReLU layer of the encoder is concatenated to the input layers of the decoder. This has been proven to enhance the reconstruction of finer features\cite{UNet}. Finally, the network produces the output fields with 2 channels (or 6 channels) for the real and imaginary part of the $H_y$ field (or $H_y$, $E_x$, $E_z$ fields). Detailed schematic of UNet structure is shown in the Supplementary Section 1.

\subsection{Dataset preparation}
During the data generation, the FDFD-based electromagnetic simulator, Ceviche \cite{ceviche}, is utilized. The perfectly matched layer (PML) \cite{PML} is adopted in the z direction and the periodic boundary condition is adopted in the x direction. The grating device is placed at the center of the simulation domain leaving at least a wavelength away from the PML to ensure the absorbing performance of it. The portion of the lower half of the simulation domain is set as silicon dioxide, the upper half is set as air. The refractive index of silicon is taken from ref. \cite{green2008self}.  An infinitely large magnetic current sheet is used for field radiation to simulate plane wave source. Only the fields and grating pattern shown in the red window in Fig. S2 are saved out for WaveY-Net training. The window has size of $[256, 64]$ and contains five rows of voxel of substrate at its bottom and seven rows of voxel of air at its top. Notice that the simulation is set such that the phase of the x component of the incident electric field is zero at the lower edge of the window in forward simulation. As for the adjoint simulation, the source is set to ensure the phase of the y component of the incident electric field is zero at the center point of the upper edge of the window. A more detailed discussion is in the Supplementary Section 1.

\subsection{Training procedure}
All the models are trained for 200 epochs in batch-sizes of 32. Adaptive Moment Estimation (Adam) \cite{kingma2014adam} optimizer is used with a $l_2$-regularization coefficient of $3\times 10^{-3}$. A learning rate of $1\times 10^{-4}$ is applied, and an exponential decaying learning rate scheduler is used with $\gamma=0.98$. The convolutional layer weights are initialized by default using Kaiming Initialization\cite{he2015delving}. All the models are trained with PyTorch version 1.8.1. We adopt a train-test split ratio of $9:1$ for all training processes.

\subsection{Evaluation metrics}
Mean absolute error (MAE) is mainly used as the evaluation metric for this study. For individual device, the normalized MAE of a certain field component is calculated by first computing the $l_1$-norm of the difference between the predicted field matrix and its corresponding ground truth, and then it is normalized through dividing by the mean absolute magnitude of the ground truth field. Note that since the field is complex (2 channel), the normalized MAE is first computed for both the real part and the imaginary part, and the average value between the two is taken. The normalized MAEs of $H_y$, $E_x$, and $E_z$ are all calculated in this manner. The normalized MAE of $\emph{\textbf{E}}$ is computed by taking the average between $E_x$ and $E_z$. Similarly, the normalized MAE of the full electromagnetic field is computed by taking the average between all three field components. The normalized MAE of the wave equation residue is calculated in the same way except that the $l_1$-norm is taken for the wave equation residue matrix computed from Equation S1.The normalized MAE of the entire set (Table \ref{tab:table1} and Fig. \ref{fig:fig4}) is evaluated by taking the mean value of the normalized MAE of each individual device within the dataset.

\subsection{Near-to-far-field transformation}
Given the near field profile of the electric field within the simulation window, we can take a horizontal slice of the E-field $E(x,z=z_1)$ and compute the far-field of the outgoing plane waves $c_q$  as a function of diffraction channel $q$ with the near- to far-field transformation as below (See Supplementary Section 5 for the derivation):

$c_q = \frac{1}{a} \sum_x E(x,z_1)e^{ik_{x,inc}x} e^{-iqKx} e^{ik_{z,q} z_1}$, 

where $a$ is the grating period,  $k_{x,inc}$ is the x-component of the incident wave vector,  $K$ is the spatial frequency of the grating computed as $K = \frac{2\pi}{a}$, and $k_{z,q}$ is the z-component of the outgoing wave vector of diffraction channel $q$. To calculate diffraction efficiency, the Poynting vector flux of the diffracted wave is integrated and divided by that of the incident field \cite{hugonin2021reticolo}.  

\subsection{Adjoint gradient calculation}
To calculate the adjoint gradient used in the adjoint variable method, the adjoint phase is first calculated given the forward electric field. More specifically, the electric field maps from the forward simulator are used to calculate the far-field response in the desired diffraction order, $\Tilde{\emph{\textbf{E}}}_{fwd}$, using the near- to far-field transformation. This far-field metric is then used to calculate diffraction adjoint phase, which is computed as: $
    \theta_{adj} = \mbox{angle}(\Tilde{\emph{\textbf{E}}}_{fwd}^*)$, where $\Tilde{\emph{\textbf{E}}}_{fwd}^*$ is the complex conjugate of $\Tilde{\emph{\textbf{E}}}_{fwd}$. The   voxel-wise gradient is computed as $\mbox{Re}(\emph{\textbf{E}}_{fwd} \cdot \emph{\textbf{E}}_{adj}  \cdot e^{i\theta_{adj}})$, which is a $64\times256$ matrix. It is then turned into a vector with dimension $1\times256$ by taking summation along the z-axis. It is cropped with a filtering operation where positive gradient is set to zero when the corresponding region is already the high-index material (Si in this case), and negative gradient is set to zero when the corresponding region is already the low-index material (air in this case). Lastly the magnitude is normalized by dividing by one half of the maximal magnitude. The boundary gradients are calculated according to methods described in refs. \cite{miller2012photonic, chen2020design}.


\section{Data availability}
The datasets generated during and/or analysed during the current study are available
from the corresponding author on reasonable request.

\bibliographystyle{naturemag}
\bibliography{refs}

\end{document}